\documentclass{elsart}

\usepackage{graphicx}
\usepackage{bm}
\usepackage{amssymb}
\usepackage{amsmath}
\begin{document}
\begin{frontmatter}

\title{The permutation entropy rate equals the metric entropy rate for
ergodic information sources and ergodic dynamical systems}
\author{Jos\'{e} M. Amig\'{o}}
\address{Centro de Investigaci\'{o}n Operativa, Universidad Miguel Hern\'{a}ndez.
03202 Elche, Spain.}
\author{Matthew B. Kennel, Ljupco Kocarev}
\address{Institute for Nonlinear Science, University of California, San Diego. La
Jolla, CA 92093-0402, USA.}

\begin{abstract}
Permutation entropy quantifies the diversity of possible orderings of the
values a random or deterministic system can take, as Shannon entropy
quantifies the diversity of values. We show that the metric and permutation
entropy rates---measures of new disorder per new observed value---are equal
for ergodic finite-alphabet information sources (discrete-time stationary
stochastic processes). With this result, we then prove that the same holds
for deterministic dynamical systems defined by ergodic maps on $n$%
-dimensional intervals. This result generalizes a previous one for piecewise
monotone interval maps on the real line (Bandt, Keller and Pompe,
\textquotedblleft Entropy of interval maps via
permutations\textquotedblright ,\ \textit{Nonlinearity} \textbf{15},
1595-602, (2002)), at the expense of requiring ergodicity and using a
definition of permutation entropy rate differing in the order of two limits.
The case of non-ergodic finite-alphabet sources is also studied and an
inequality developed. Finally, the equality of permutation and metric
entropy rates is extended to ergodic non-discrete information sources when
entropy is replaced by differential entropy in the usual way.
\end{abstract}


\end{frontmatter}
%\maketitle

\section{\noindent Introduction}

The entropy rate is a key parameter associated with stochastic processes,
information sources and dynamical systems. Roughly speaking, the entropy
rate quantifies the average uncertainty, disorder or irregularity generated
by a process or system per `time' unit and, it is the primary subject of
fundamental results in information and coding theory (Shannon's noiseless
coding theorem) and statistical mechanics (second law of thermodynamics). It
is not surprising, therefore, that this notion, appropriately generalized
and transformed, is ubiquitous in many fields of mathematics and science
when randomness or `random-like' behavior is at the heart of the theory or
model being studied.

For definiteness consider a stationary information source emitting a
time-series of observed values $x_{1},\ldots ,x_{n}$ in a continuous state
space ---formally, draws from the random variables $X_{1},\ldots ,X_{n}$.
Since the realization of a non-discrete random variable cannot be observed
exactly (this would mean an infinite amount of information), the observer
has to content himself with a finite degree of accuracy. Generally speaking,
the metric or Shannon entropy rate of an information source is the rate of
new information it generates per unit time (as the metric or
Kolmogorov-Sinai entropy rate of a deterministic dynamical system is a
measure of its pseudo-randomness or chaotic behavior). Given a certain
discretization scale $\Delta $ of the state space, the metric (Shannon)
entropy rate $h_{m}$ of the discretized information source $\mathbf{X}%
^{\Delta }=(X_{n}^{\Delta })_{n\in \mathbb{N}}$ is 
\begin{equation*}
h_{m}(\mathbf{X}^{\Delta })=\lim_{L\rightarrow \infty }\frac{1}{L}%
H_{m}\left( X^{\Delta }{}_{1}^{L}\right) ,
\end{equation*}%
with $X^{\Delta }{}_{1}^{L}=X_{1}^{\Delta }\ldots X_{L}^{\Delta }$ a length $%
L$ word of symbols $X^{\Delta }$ discretized at resolution $\Delta $ from $%
X_{1}^{L}=X_{1}\ldots X_{L}$. We use $H_{m}(Z)$ for the entropy of the
discrete random variable $Z$, i.e., $H_{m}(Z)\equiv H_{m}(\Pr
(Z))=-\sum_{z}\Pr (z)\log _{2}\Pr (z)$ for the probability distribution $\Pr
(z)$ of $Z$. We come back to the metric entropy and entropy rate in the next
section, where we set the conceptual background of this paper on a more
formal footing.

Consider a length $L$ word of observables $X^{\Delta }{}_{1}^{L}$. Assuming
there exists a natural order relation on the state space of the source $%
\mathbf{X}^{\Delta }$ (e.g., real scalars or vectors with a defined
lexicographic ordering), each block of observations $X^{\Delta }{}_{1}^{L}$
selects one particular \emph{permutation} $\Pi $ out of the $L!$ possible
permutations. For example, if $X_{2}^{\Delta }\leq X_{1}^{\Delta }\leq
X_{3}^{\Delta }$, then the corresponding permutation can be expressed
explicitly as $\Pi (X^{\Delta }{}_{1}^{3})=(2,1,3)$. Note that the mapping
from $X^{\Delta }$-orderings to permutations can be many-to-one when there
are repeated values; to overcome this shortcoming, we will use `ranks' below
(see Sect. 3), so that words defining the same permutation have the same
rank variables which, in turn, can be identified with the corresponding
permutation. Bandt and Pompe \cite{Bandt2} defined the \emph{\ permutation
entropy of order} $L$ as 
\footnote{The factor $1/(L-1)$ is used instead $1/L$ because $\Pi (X^{\Delta
}{}_{1}^{1})=1$ contributes nothing to the entropy. This choice is, of
course, inconsequential when $L\rightarrow \infty $, but it is preferable
for numerical simulations and the applications we discuss in the last
section.} 
\begin{equation*}
\bar{H}_{m}^{\ast }(X^{\Delta }{}_{1}^{L})=\frac{1}{L-1}H_{m}(\Pi (X^{\Delta
}{}_{1}^{L})),
\end{equation*}%
with $\Pr \ \Pi (X^{\Delta }{}_{1}^{L})$ being the probability of observing
any particular permutation given a block of observables. In direct analogy
to the Shannon entropy rate, the \emph{permutation entropy rate} at
resolution $\Delta $ is hence defined as (following the notation of \cite%
{Bandt}) 
\begin{equation}
h_{m}^{\ast }(\mathbf{X}^{\Delta }):=\lim_{L\rightarrow \infty }\bar{H}%
_{m}^{\ast }(X^{\Delta }{}_{1}^{L}).  \label{hperemut}
\end{equation}

For \textit{deterministic} maps $f$ of a proper interval $I\subset
\mathbb{R} $ with a finite number of monotony segments, Bandt, Keller
and Pompe \cite{Bandt,Bandt2} analytically and numerically
investigated a permutation entropy rate we denote by $h_{m}^{\ast
\mathrm{BKP}}(f)$, based on the entropy of certain partitions, proving
that it exists and, in fact, equals the metric (Kolmogorov-Sinai)
entropy rate $h_{m}(f)$. They also prove this equality for the
topological versions of permutation and ordinary entropy
rates. Relative changes in $h_{m}^{\ast \mathrm{BKP}}$ estimated numerically
from time-series from the logistic map tended to track very well, over
a wide range of varying nonlinearity parameter, the behavior of
$h_{m}$ (estimated from the positive Lyapunov exponent of the map
directly). There remained a substantial bias, though it was nearly
constant over parameters.

The correspondence observed in \cite{Bandt2} between permutation entropy and
metric entropy rates of time series is not coincidental, nor restricted to
one-dimensional dynamics. Under only the assumption of ergodicity, we show
that the permutation entropy rate of stationary, finite-alphabet random
processes equals the metric entropy rate. A similar result follows for the
permutation and metric \textit{differential} entropy rates of non-discrete
sources. With these results on stochastic processes in the hand, we further
show that for \textit{ergodic} maps on $d$-dimensional intervals $I^{d}$ the
two entropy rates are also equal. In doing so, we define the permutation
entropy rate as $h_{m}^{\ast }(f)=\lim_{\Delta \rightarrow 0}h_{m}^{\ast }(%
\mathbf{X}^{\Delta })$, where $\mathbf{X}^{\Delta }$ stands now for the
`simple observations' of $f$ supplied by a discretization of $I^{d}$ with
resolution $\Delta $ ---a finite-state stochastic process. The generality of
all these results gives a strong support to our approach, which provides a
unified treatment for stochastic and deterministic dynamical systems.

This paper is organized as follows. For the reader's convenience we review
in Sect. II the theoretical background and fix the notation. Sect. III
contains one of the main results of this paper, namely, $h_{m}=h_{m}^{\ast }$
for \textit{ergodic} finite-alphabet stochastic processes (Theorem 1). This
result is generalized in Sect. IV to non-discrete ergodic information
sources using the differential entropy rate (Theorem 2) and, in Sect. V, to
maps on $d$-dimensional intervals (Theorem 3). We also mention in Sect. III
that $h_{m}^{\ast }\geq h_{m}$ for \textit{non-ergodic} finite-alphabet
sources; the proof can be found in Appendix B. Sect. V contains the main
result on finite-dimensional maps, and Sect. VI, a discussion of the two
definitions of permutation entropy. Finally, in Sect. VII we show some
numerical examples and discuss open practical issues in using permutation
entropies in time-series analysis.

\section{\noindent Theoretical framework}

\subsection{\noindent Stochastic processes and dynamical systems\textbf{\ }}

Let $\mathbb{R}^{d\mathbb{N}}=\{x=(x_{n})_{_{n\in \mathbb{N}}}:x_{n}\in 
\mathbb{R}^{d}\}$, $\mathcal{B}$ the product sigma-algebra of $\mathbb{R}^{d%
\mathbb{N}}$ generated by the Borel sets of $\mathbb{R}^{d}$, and $\sigma $
the (left) shift transformation on $\mathbb{R}^{d\mathbb{N}}$, $(\sigma
x)_{n}=x_{n+1}$. Let $(\Omega ,\mathcal{F},\mu )$ be a probability space,
i.e., $\Omega $ is a nonempty set, $\mathcal{F}$ is a sigma-algebra of
subsets of $\Omega $ and $\mu $ is a (positive) measure on $(\Omega ,%
\mathcal{F})$. Any \textit{stationary} stochastic (or random) process in
discrete time $\mathbf{X}=(X_{n})_{n\in \mathbb{N}}$ on the probability
space $(\Omega ,\mathcal{F},\mu )$ with values in $\mathbb{R}^{d}$
corresponds in a standard way to the shift dynamical system $(\mathbb{R}^{d%
\mathbb{N}},\mathcal{B},m,\sigma )$ via the map $\phi :\Omega \rightarrow 
\mathbb{R}^{d\mathbb{N}}$ defined by $(\phi \omega )_{n}=X_{n}(\omega )$, $%
n\in \mathbb{N}$. The probability measure $m$ is defined on the Borel sets $B
$ of $\mathbb{R}^{d\mathbb{N}}$ by%
\begin{equation*}
m(B):=\mu (\phi ^{-1}B)
\end{equation*}%
($\phi ^{-1}B\in \mathcal{F}$ because $X_{n}$ is $\mathcal{F}$-measurable
for all $n$) and it is $\sigma $-invariant (i.e., $m\circ \sigma ^{-1}=m$)
because of the stationarity of $\mathbf{X}$. The measure $m$ is sometimes
called the induced probability measure or distribution on the space of
possible outputs of the random process. Moreover, if $\pi _{n}:\mathbb{R}^{d%
\mathbb{N}}\rightarrow \mathbb{R}^{d}$ is the projection onto the $n$th
component, $\pi _{n}x=x_{n}=X_{n}(\omega )$ (or $\pi _{n}=X_{n}\circ \phi
^{-1}$), then the `sampling function' $\mathbf{\pi }=(\pi _{n})$ has the
same joint distributions on $\mathbb{R}^{d\mathbb{N}}$ as $\mathbf{X}=(X_{n})
$ on $\Omega $, i.e., both processes are equivalent. Any point $x$ of the 
\textit{state space} $\mathbb{R}^{d\mathbb{N}}$ is a possible realization
(or `sample path') of the whole process. Such one-sided random processes
provide better models than the two-sided processes $(X_{n})_{n\in \mathbb{Z}}
$ for physical information sources that must be turned on at some time and
thus we will use both denominations interchangeably in this paper.

We will also refer to the shift dynamical system $(\mathbb{R}^{d\mathbb{N}},%
\mathcal{B},m,\sigma )$ as the (sequence space) \textit{model} of the
stochastic process or information source $\mathbf{X}$. Sometimes $\mathbb{Z}%
_{+}=\{0,1,...\}$ is used instead of $\mathbb{N}$ to number the random
variables $X_{n}$ and their samples $x_{n}$ (we do so in Sect. 4). Models
allow to focus on the random process itself as given by the probability
distribution on their outputs, dispensing with a perhaps complicated
underlying probability space. As usual, we will also identify $X_{n}$ with $%
\pi _{n}=\pi _{0}\circ \sigma ^{n}.$

Finite-state or finite-alphabet sources $\mathbf{S}=(S_{n})_{n\in \mathbb{N}}
$ on $(\Omega ,\mathcal{F},\mu )$, where $S_{n}:\Omega \rightarrow A$ with
alphabet $A=\{a_{1},\ldots ,a_{\left\vert A\right\vert }\}$, are dealt with
in a similar way to the previous, non-discrete sources and, as a matter of
fact, most of the general setup, properties and observations above apply 
\textit{mutatis mutandis} to this simpler case. The sequence space of the
corresponding model is now $A^{\mathbb{N}}=\{s=(s_{n})_{_{n\in \mathbb{N}%
}}:s_{n}\in A\}$, $A$ being endowed with the discrete topology; let $%
\mathcal{Z}$ be the product sigma-algebra of $A^{\mathbb{N}}$ generated by
the elements of $A$. Since no confusion will arise, we continue denoting by $%
\sigma $ the shift on $A^{\mathbb{N}}$, $(\sigma s)_{n}=s_{n+1}$, and by $m$
the $\sigma $-invariant measure on $(A^{\mathbb{N}},\mathcal{Z})$ defined as
the pushforward of $\mu $ by the map $\phi :\Omega \rightarrow A^{\mathbb{N}}
$, $(\phi \omega )_{n}=S_{n}(\omega )$. The finite order probability
distribution of $\mathbf{S}$, $\Pr (S_{i_{1}}=s_{i_{1}},\ldots
,S_{i_{n}}=s_{i_{n}})=:\Pr (s_{i_{1}},\ldots ,s_{i_{n}})$, can be
alternatively expressed by means of the probability distribution on the
outputs of $\mathbf{S}$, 
\begin{eqnarray}
&&\Pr (s_{i_{1}},\ldots ,s_{i_{n}})  \notag \\
&=&\mu \left\{ \omega \in \Omega :S_{i_{1}}(\omega )=s_{i_{1}},\ldots
,S_{i_{n}}(\omega )=s_{i_{n}}\right\}   \notag \\
&=&m\left\{ \xi \in A^{\mathbb{N}}:\xi _{i_{1}}=s_{i_{1}},\ldots ,\xi
_{i_{n}}=s_{i_{n}}\right\}   \label{P=p}
\end{eqnarray}%
for any $i_{1},\ldots ,i_{n}\in \mathbb{N}$ and $s_{i_{1}},\ldots
,s_{i_{n}}\in A$.

In this paper we will consider mostly finite-alphabet sources, although
these will also occasionally arise as discretizations or quantizations $%
\mathbf{X}{^{\Delta }}$ of sources $\mathbf{X}$ taking values on a proper
interval $I^{d}$ of $\mathbb{R}^{d}$ ($I^{d}\varsubsetneq \mathbb{R}^{d}$ in
symbols) endowed with Lebesgue measure $\lambda $. Formally, this means that
there exists a (usually, uniform) partition $\delta =\{\Delta
_{1},...,\Delta _{\left\vert \delta \right\vert }\}$ of $I^{d}$ into a
finite number of $\lambda $-measurable subsets such that $X_{n}^{\Delta }$
is the discrete random variable defined by%
\begin{eqnarray*}
\Pr \left( X_{n}^{\Delta }=i\right)  &=&\mu \left\{ \omega \in \Omega
:X_{n}^{\Delta }(\omega )\in \Delta _{i}\right\}  \\
&=&m\left\{ \xi \in A^{\Delta \mathbb{N}}:\xi _{n}=i\right\} =\int_{\Delta
_{i}}dF(x),
\end{eqnarray*}%
where $F(x)=\Pr (X_{n}^{\Delta }\leq x)=\mu \{\omega \in \Omega
:X_{n}^{\Delta }(\omega )\leq x\}$ is the common distribution function to
all $X_{n}^{\Delta }$ (in case $X_{n}^{\Delta }$ is a vector random
variable, the inequality is understood component-wise), $m$ is the induced
probability measure on the outputs and $A^{\Delta }=\{1,...,\left\vert
\delta \right\vert \}$ is the alphabet of $\mathbf{X}{^{\Delta }}$. If $%
X_{n}^{\Delta }$ has a density function $\rho (x)$ (formally, the
Radon-Nykodim derivative of $F$ with respect to $\lambda $), then $\Pr
\left( X_{n}^{\Delta }=i\right) =\int_{\Delta _{i}}\rho (x)dx$. Distribution
functions and densities of higher finite order are analogously defined. For $%
\Delta $, the `discretization scale'\ or `resolution'\ we referred to in the
Introduction, one can take any measure of the `coarseness' of $\delta $,
say, the largest diameter of its elements, also called the norm of $\delta $%
, $\left\Vert \delta \right\Vert $.

\subsection{\noindent Entropy rate of dynamical systems and stochastic
processes}

Let $(\Omega ,\mathcal{F},\mu )$ be a probability space and $f:\Omega
\rightarrow \Omega $ a $\mu $-preserving transformation, i.e., $\mu
(f^{-1}A)=\mu (A)$ for all $A\in \mathcal{F}$. Given the dynamical system $%
(\Omega ,\mathcal{F},\mu ,f)$ and a finite partition $\alpha
=\{A_{1},...,A_{\left\vert \alpha \right\vert }\}\subset \mathcal{F}$ of $%
\Omega $, the entropy of $f$ with respect to $\alpha $ is defined as 
\begin{equation}
h_{\mu }(f,\alpha ):=\lim_{L\rightarrow \infty }\frac{1}{L}H_{\mu }\left(
\bigvee\limits_{i=0}^{L-1}f^{-i}\alpha \right) ,  \label{hmu}
\end{equation}%
where $\vee _{i=0}^{L-1}f^{-i}\alpha =\{\cap _{i=0}^{L-1}f^{-i}A_{j_{i}}\}$
is the least common refinement of the partitions $\left\{ \alpha
,f^{-1}\alpha ,...,f^{-L+1}\alpha \right\} $ and $H_{\mu }(\beta
):=-\sum_{j=1}^{\left\vert \beta \right\vert }\mu (B_{j})\log \mu (B_{j})$
for any finite partition $\beta =\{B_{1},...,B_{\left\vert \beta \right\vert
}\}\subset \mathcal{F}$. The metric or Kolmogorov-Sinai entropy rate of map $%
f$ is then defined as: 
\begin{equation}
h_{\mu }(f):=\sup_{\alpha }\,h_{\mu }(f,\alpha ).
\end{equation}%
The convergence in (\ref{hmu}) can be proved to be monotonically decreasing 
\cite{Katok}. Assuming logarithms base 2 everywhere herein, $h_{\mu }(f)$
has units of bits per symbol or time unit, if $n$ is interpreted as discrete
time. By convention, $0\cdot \log 0:=\lim_{x\rightarrow 0+}x\log x=0$. In an
information-theoretical setting, $h_{\mu }(f,\alpha )$ represents the
long-term average of the information gained per unit time with respect to a
certain partition and $h_{\mu }(f)$ the maximum information per unit time
available from any stationary process generated by the source, typically
equal to the sum of the positive Lyapunov exponents by the Pesin theorem. If
there exists finite $\gamma $ such that $h_{\mu }(f,\gamma )=h_{\mu }(f)$,
then $\gamma $ is called a generator, or generating partition, of $f$.

Given a discrete alphabet source $\mathbf{S}=(S_{n})$ with model $(A^{%
\mathbb{N}},\mathcal{Z},m,\sigma )$, the (Shannon) entropy of the random
variables $S_{1}^{L}:=S_{1}\ldots S_{L}$ is%
\begin{equation*}
H_{m}(S_{1}^{L}):=H_{m}\left( \bigvee\limits_{i=0}^{L-1}\sigma ^{-i}\zeta
\right) ,
\end{equation*}%
where $\zeta $ $=\{C_{1},...,C_{\left\vert A\right\vert }\}$ is the
partition of $A^{\mathbb{N}}$ consisting of the basic `cylinder sets' $%
C_{i}=\{s\in A^{\mathbb{N}}:s_{1}=a_{i}\}$, $1\leq i\leq \left\vert
A\right\vert $. According to (\ref{P=p}),%
\begin{equation*}
H_{m}(S_{1}^{L})=-\sum \Pr (s_{1},\ldots ,s_{L})\log \Pr (s_{1},\ldots
,s_{L}),
\end{equation*}%
and, correspondingly, the \textit{entropy rate} (or uncertainty) of the
source is defined as $h_{m}(\mathbf{S}):=h_{m}(\sigma ,\zeta )=h_{m}(\sigma
) $ since $\zeta $ is a (one-sided) generator of $\mathcal{Z}$, i.e., 
\begin{equation*}
h_{m}(\mathbf{S})=\lim_{n\rightarrow \infty }\frac{1}{L}H_{m}(S_{1}^{L}).
\end{equation*}%
In other words, the Shannon entropy rate of $\mathbf{S}$ is, by definition,
the Kolmogorov-Sinai entropy rate of its sequence space model. This explains
our using `metric' to refer to both concepts, independently of the random or
deterministic nature of the system considered. Sometimes we will also use
the $n$\textit{th order entropy} of $\mathbf{S}$, 
\begin{equation*}
\bar{H}_{m}(S_{1}^{L}):=\frac{1}{L}H_{m}(S_{1}^{L}),
\end{equation*}%
so that $h_{m}(\mathbf{S})=\lim_{L\rightarrow \infty }\bar{H}_{m}(S_{1}^{L})$%
. In general, $S_{i}^{j}$ stands for the string $S_{i}\ldots S_{j}$.

Other dynamical, statistical or information-theoretical concepts like
conditional entropy, mutual information, ergodicity, mixing properties,
etc., are also defined via the sequence space model. For example, $\mathbf{S}
$ is said to be \textit{ergodic} if $(A^{\mathbb{N}},\mathcal{Z},m,\sigma )$
is ergodic, i.e., for $C_{1},C_{2}\in \mathcal{Z}$ with $m(C_{1})>0$, $%
m(C_{2})>0$, there exists $n>0$ such that $m(C_{1}\cap \sigma ^{-n}C_{2})>0$.

If, more generally, $\mathbf{X}$ is a non-discrete scalar or vector source
with outcomes on an interval $I^{d}\varsubsetneq \mathbb{R}^{d}$, define its 
\textit{differential entropy rate} as%
\begin{equation}
h_{m}(\mathbf{X}):=\lim_{\Delta \rightarrow 0}\left( h_{m}(\mathbf{X}%
^{\Delta })+\log \Delta \right) ,  \label{diffentropy}
\end{equation}%
where $\mathbf{X}^{\Delta }$ is a uniform discretization of $\mathbf{X}$
with resolution scale $\Delta $. The differential entropy shows how the
average rate of information furnished by a quantization of resolution $%
\Delta $ differs from $\left\vert \log \Delta \right\vert $ when $\Delta
\rightarrow 0$. If $X^{\Delta }{}_{1}^{L}$ happens to have a density
function $\rho (x_{1},...,x_{L})${\ for every }$L\geq 1$, then 
\begin{equation*}
h_{m}(\mathbf{X})=\int_{I^{d}}\rho (x_{1},...,x_{L})\log \rho
(x_{1},...,x_{L})d^{L}x.
\end{equation*}

\section{\noindent Permutations and the metric entropy rate of
finite-alphabet sources}

\noindent Given a finite-alphabet source $\mathbf{S}=(S_{n})$ with model $%
(A^{\mathbb{N}},\mathcal{Z},m,\sigma )$, each possible permutation of a
block of length $L$, e.g., $S_{1}^{L}:=S_{1}\ldots S_{L}$, can be indexed as
a word of \emph{ranks}, each an integer in successively larger alphabets. In
particular, define for $n\geq 1$ the rank variable $R_{n}=\left\vert
\{S_{i},1\leq i\leq n:S_{i}\leq S_{n}\}\right\vert =\sum_{i=1}^{n}\delta
(S_{i}\leq S_{n})$, where, as usual, the $\delta $-function of a proposition
is $1$ if it holds and $0$ otherwise. By definition, $R_{n}$ is a \textit{\
discrete} random variable on $\Omega $ with range $\{1,\ldots ,n\}$ and the
sequence $\mathbf{R}=(R_{n})$ builds a discrete-time non-stationary process.
Then the permutation $\Pi (S_{1}^{L})$ in (\ref{hperemut}) can also be
viewed as the word $R_{1}^{L}=R_{1}\ldots R_{L}$, the relation between both
being one-to-one. The many-to-one relation between $S_{1}^{L}$ and $%
R_{1}^{L} $ is written as $R_{1}^{L}=\varphi (S_{1}^{L})$.

For example, consider a source $\mathbf{S}$ over the alphabet $\{1,2,3\}$.
Suppose we observe the word $S_{1}^{3}=1,3,3$. Then, $R_{1}^{3}=\varphi
(S_{1}^{3})=1,2,3$, (of course other strings, e.g., $1,1,1$ or $2,2,2$, also
map to $R_{1}^{3}=1,2,3$) and $\Pi (S_{1}^{3})=(1,2,3)$. The string $1,3,3$
could be counted as matching both the ordering $S_{1}\leq S_{2}\leq S_{3}$
and $S_{1}\leq S_{3}\leq S_{2}$. By using ranks, by contrast, the measure
associated with each word is unambiguously associated with one permutation,
and the rest of our development follows this approach.

The \textit{permutation entropy rate} of $\mathbf{S}$ is then defined as 
\begin{equation*}
h_{m}^{\ast }(\mathbf{S}):=\lim_{L\rightarrow \infty }\bar{H}_{m}\left(
R_{1}^{L}\right) ,
\end{equation*}%
alternatively to the definition (\ref{hperemut}), with 
\begin{eqnarray*}
\bar{H}_{m}^{\ast }\left( S_{1}^{L}\right) &=&\bar{H}_{m}\left(
R_{1}^{L}\right) \\
&=&-\frac{1}{L-1}\sum \Pr (r_{1},...,r_{L})\log \Pr (r_{1},...,r_{L})
\end{eqnarray*}%
defined to be the \textit{permutation entropy of order} $L\geq 2$ of $%
\mathbf{S}$. Remember that the overbar notation $\bar{H}$ means that the
relevant factor of $1/L$ or $1/(L-1)$ has been included for the entropy of a
block of length $L$.

Let $\sigma _{L}$ denote the set of permutations of $\{1,...,L\}$ for the
time being. We say that the word $S_{1}^{L}$ is of type $\pi \in \sigma _{L}$
if $R_{1}^{L}=\varphi (S_{1}^{L})$ defines the permutation $\pi $. It
follows $s_{\pi (1)}\leq \ldots \leq s_{\pi (L)}$. The cylinder sets 
\begin{equation*}
C_{\pi }:=\{s\in A^{\mathbb{N}}:s_{1}^{L}\text{ is of type }\pi \}
\end{equation*}%
such that $C_{\pi }\neq \varnothing $ build a partition of $A^{\mathbb{N}}$
with $m(C_{\pi })=\Pr (R_{1}^{L}=r_{1}^{L})$, $1\leq r_{k}\leq k$ for $%
k=1,\ldots ,L$. Therefore%
\begin{equation}
\bar{H}_{m}^{\ast }\left( S_{1}^{L}\right) =-\frac{1}{L-1}\sum_{\pi \in
\sigma _{L}}m(C_{\pi })\log m(C_{\pi }).  \label{Qpi}
\end{equation}%
That is, the permutation entropy is sensitive to the measures of
non-trivial order relationships observed in a word, as the Shannon entropy
is sensitive to the measures of the different word values themselves.

%remark1
Observe as a technical point for later
reference that, if 
\begin{equation*}
Q_{\pi }:=\{s\in A^{\mathbb{N}}:s_{\pi (1)}\leq s_{\pi (2)}\leq ...\leq
s_{\pi (L)}\},
\end{equation*}%
then $C_{\pi }\varsubsetneq Q_{\pi }$ due to words $s_{1}^{L}$ with repeated
letters: if $s_{i}\neq s_{j}$ for every $1\leq i,j\leq L$, then $s\in C_{\pi
}$ if and only if $s\in Q_{\pi }$.

%\noindent \textbf{Lemma 1.}
\begin{lem} \label{lemma1}
Given an ergodic information source $\mathbf{S}$, 
\begin{equation*}
\lim_{k\rightarrow \infty }H_{m}(R_{k+1}^{k+l}|S_{1}^{k})=\lim_{k\rightarrow
\infty }H_{m}(S_{k+1}^{k+l}|S_{1}^{k})
\end{equation*}%
for all $l\geq 1$.
\end{lem}

That is, given a sufficiently long tail of previously observed symbols, the
later ranks can be predicted virtually as well as the symbols themselves.
Heuristically, this is because the distribution of rank variable $R_{k+1}$
for $k$ sufficiently large depends effectively on only the cumulative
distribution function of the source, approximated by the normalized sum of $%
S_{1}^{k}$. In turn this means that the information contained in $R_{k+1}$
is the same as the information in $S_{k+1}$. The proof, and an elementary
example, is given in Appendix~\ref{sec:appendix1}. With Lemma~\ref{lemma1} in hand, we
turn to our first main result, the equality between permutation and metric
entropy for finite-alphabet stochastic processes.

\begin{thm} \label{theorem:finite-alphabet} For finite-alphabet ergodic sources 
$\mathbf{S}$ the permutation entropy rate exists and equals the metric
entropy rate: $h_{m}^{\ast }(\mathbf{S})=h_{m}(\mathbf{S})$.
\end{thm}
\begin{pf} We prove inequalities in both directions.\\
(a) $\lim \, \sup_{L\rightarrow \infty }\bar{H}_{m}^{\ast }(S_{1}^{L})\leq
h_{m}(\mathbf{S})$. \noindent Given $S_{1}^{L}$, the corresponding rank
variables are uniquely determined via $R_{1}^{L}=\varphi (S_{1}^{L})$. By~%
\cite{Cover} (Ch 2, exercise 5), $H(\varphi (Z)) \leq H(Z)$ for any discrete
random variable $Z$, so $H_{m}(R_{1}^{L})\leq H_{m}(S_{1}^{L})$ and thus $%
\lim \sup_{L\rightarrow \infty }\bar{H}_{m}(R_{1}^{L})\leq \lim
\sup_{L\rightarrow \infty }\bar{H}_{m}(S_{1}^{L})=h_{m}(\mathbf{S})$.

(b) $\lim \, \inf_{L\rightarrow \infty }\bar{H}_{m}^{\ast }(S_{1}^{L})\geq
h_{m}(\mathbf{S})$. There are several ways to prove this inequality.
Consider, for instance, 
\begin{eqnarray*}
\lim_{L\rightarrow \infty }\inf \bar{H}_{m}^{\ast
}(S_{1}^{L})&=&\lim_{L\rightarrow \infty }\inf \frac{1}{L}H_{m}(R_{1}^{L}) \\
&=&\lim_{L\rightarrow \infty }\inf \frac{1}{L}\left[
H_{m}(R_{L}|R_{1}^{L-1})+\ldots +H_{m}(R_{L^{\ast }+1}|R_{1}^{L^{\ast
}})+H_{m}(R_{1}^{L^{\ast }})\right]
\end{eqnarray*}%
for any $L^{\ast }<L$, where we have applied the chain rule for entropy. As $%
R_{1}^{k}=\varphi (S_{1}^{k})$ we apply the data processing inequality $%
H(Y|\varphi (Z))\geq H(Y|Z)$ \cite{Cover} to all elements of the first term
on the rhs: 
\begin{equation*}
\lim_{L\rightarrow \infty }\inf \bar{H}_{m}^{\ast }(S_{1}^{L}) \geq \lim_{L\rightarrow \infty }\inf \frac{1}{L}\left[ H_{m}(R_{L}|S_{1}^{L-1})+
\ldots +H_{m}(R_{L^{\ast }+1}|S_{1}^{L^{\ast }})+H_{m}(R_{1}^{L^{\ast }}) 
\right].
\end{equation*}
By Lemma~\ref{lemma1}, for any $\varepsilon >0$ there is some $L^{\ast }$ such that $%
\left\vert H_{m}(S_{L}|S_{1}^{L-1})-H_{m}(R_{L}|S_{1}^{L-1})\right\vert
<\varepsilon $ for $L>L^{\ast }$, so 
\begin{eqnarray*}
\lim_{L\rightarrow \infty }\inf \bar{H}_{m}^{\ast }(S_{1}^{L}) 
&\geq&\lim_{L\rightarrow \infty }\inf \left( \frac{1}{L}\left[H_{m}(S_{L}|S_{1}^{L-1})+\ldots  +H_{m}(S_{2}|S_{1})+H_{m}(S_{1})\right] \right. \\
&&\,\,\,\left. +\frac{1}{L}\left[ H_{m}(R_{1}^{L^{\ast
}})-H_{m}(S_{1}^{L^{\ast }})\right] -\left( \frac{L-L^{\ast }}{L}\right)
\varepsilon \right) \\
&=&h_{m}(\mathbf{S})-\varepsilon .
\end{eqnarray*}%
The existence of the limit and equality follows from (a) and (b).\qed
\end{pf}

More generally, we can only show an inequality for \textit{%
non-ergodic} cases, namely,%
\begin{equation}
\lim_{L\rightarrow \infty }\inf \bar{H}_{m}^{\ast }(S_{1}^{L})\geq h_{m}(%
\mathbf{S}).  \label{Nonergodic}
\end{equation}%
The proof of (\ref{Nonergodic}) uses the ergodic decomposition of the
entropy rate and is given in Appendix B.

\section{Non-discrete information sources}

Information sources can have also non-discrete alphabets, although
their outcomes are only observable with a finite precision. In this
case, it is well-known that Shannon's entropy rate, defined as the
limit over ever finer uniform quantizations of the source, diverges
logarithmically with the quantization scale. In order to obtain a
finite measure of the asymptotic behavior of such quantizations, one
has to resort to the differential entropy rate (\ref{diffentropy})
instead. It turns out that Theorem~\ref{theorem:finite-alphabet} can
be extended to scalar and vector ergodic non-discrete sources if
entropy is replaced by differential entropy.

Let $\mathbf{X}=(X_{n})$ be a scalar or vector ergodic source taking values
on an interval $I^{d}\varsubsetneq \mathbb{R}^{d}$, $d\geq 1$. In case $d>1$
(vector sources), $I^{d}$ is supposed to be endowed with the product (or
lexicographical) order: $x\leq x^{\prime }$ if $x_{k}=x_{k}^{\prime }$ for $%
k=d,d-1,...,d-s>1$ and $x_{d-s-1}<x_{d-s-1}^{\prime }$ (other conventions
are also possible). With the equality between permutation and metric entropy
rates for ergodic finite-alphabet sources, we now consider the source $%
\mathbf{X}$ \textit{uniformly} discretized to an alphabet $A^{\Delta
}=\{1,\ldots ,N\}$ by means of a partition $\delta =\{\Delta _{1},...,\Delta
_{N}\}$ of $I^{d}$ with $\lambda (\Delta _{i})=\lambda (I^{d})/N=:\Delta $
for $1\leq i\leq N$, where $\lambda $ is, as before, Lebesgue measure. One
can then define the ranks $R_{n}^{\Delta }:\Omega \rightarrow \{1,\ldots
,n\} $ of blocks of discretized symbols $X^{\Delta }{}_{1}^{L}$ in the known
way: $R_{n}^{\Delta }=\sum_{i=1}^{n}\delta (X_{i}^{\Delta }\leq
X_{n}^{\Delta })$, $1\leq n\leq L$. If $(A^{\Delta \mathbb{N}},\mathcal{Z}%
^{\Delta },m^{\Delta },\sigma )$ is the sequence space model for $\mathbf{X}%
^{\Delta } $, we define the permutation entropy rate at resolution $\Delta $
as usual: $h_{m^{\Delta }}^{\ast }(\mathbf{X}^{\Delta }):=\lim_{L\rightarrow
\infty }\bar{H}_{m^{\Delta }}(R{^{\Delta }}_{1}^{L})$. We can take now the
limit $\Delta \rightarrow 0$ and, analogously to (\ref{diffentropy}),
define the \textit{differential permutation entropy rate} of
$\mathbf{X}$ as,
\begin{eqnarray*}
h_{m}^{\ast }(\mathbf{X}) &:=&\lim_{\Delta \rightarrow 0}\left( h_{m^{\Delta
}}^{\ast }(\mathbf{X}^{\Delta })+\log \Delta \right) \\
&=&\lim_{\Delta \rightarrow 0}\left( \lim_{L\rightarrow \infty }\bar{H}%
_{m^{\Delta }}(R{^{\Delta }}_{1}^{L})+\log \Delta \right) .
\end{eqnarray*}%
This yields:
\begin{thm} \label{theorem:non-discrete} Suppose $\mathbf{X}$ is an
ergodic non-discrete source. Then $h_{m}^{\ast }(\mathbf{X})=h_{m}(\mathbf{X%
})$, that is, the differential permutation and metric entropy rates of 
$\mathbf{X}$ are equal.
\end{thm}
\begin{pf}
If $(\mathbb{R}^{d\mathbb{N}},\mathcal{B},m,\sigma
)$ is ergodic, so is $(A^{\Delta \mathbb{N}},\mathcal{Z}^{\Delta },m^{\Delta
},\sigma )$. By Theorem~\ref{theorem:finite-alphabet}, $h_{m^{\Delta }}^{\ast }(\mathbf{X}^{\Delta
})=h_{m^{\Delta }}(\mathbf{X}^{\Delta })$, so 
\begin{equation*}
h_{m}^{\ast }(\mathbf{X})=\lim_{\Delta \rightarrow 0}\left( h_{m^{\Delta }}(%
\mathbf{X}^{\Delta })+\log \Delta \right) =h_{m}(\mathbf{X}),
\end{equation*}%
where $h_{m}(\mathbf{X})$ is the metric differential entropy rate of $
\mathbf{X}$.\qed
\end{pf}

\section{\noindent Permutations and the metric entropy of ergodic maps}

In this section we will use our result on finite-alphabet stochastic
processes to show that the equality between permutation and Kolmogorov-Sinai
entropy rate applies to ergodic maps on finite-dimensional intervals.

Let $I^{d}$\ be a proper interval of $\mathbb{R}^{d}$ endowed with the
sigma-algebra $\left. \mathcal{B}\right\vert _{I^{d}}=\mathcal{B}\cap I^{d}$%
, the restriction of Borel sigma-algebra of $\mathbb{R}^{d}$ to $I^{d}$, and
let $f:I^{d}\rightarrow I^{d}$ be a $\mu $-preserving transformation, with $%
\mu $ being a measure on $(I^{d},\left. \mathcal{B}\right\vert _{I^{d}})$.
In order to define the permutation entropy of $f$, we consider first product
partitions 
\begin{equation*}
\iota =\prod_{k=1}^{d}\{I_{1,k},\ldots ,I_{N_{k},k}\}
\end{equation*}%
of $I^{d}$ into $N^{d}:=N_{1}...N_{d}$ subintervals of lengths $\Delta
_{j,k} $, $1\leq j\leq N_{k}$, in each coordinate $k$, defining $\left\Vert
\iota \right\Vert =\max_{j,k}\Delta _{j,k}$. The intervals are
lexicographically ordered in each dimension, i.e., points in $I_{j,k}$ are
smaller than points in $I_{j+1,k}$ and for the multiple dimensions a
lexicographic order is defined, $I_{j,k}<I_{j,k+1}$, so there is an order
relation between all the $N^{d}$ partition elements, and we can enumerate
them with a single index $i\in \lbrack 1,N^{d}]$:%
\begin{equation*}
\iota =\{I_{i}^{d}:1\leq i\leq N^{d}\}\text{, \ }I_{i}^{d}<I_{i+1}^{d}
\end{equation*}

Next define a collection of \textit{simple observations} $\mathbf{S}^{\iota
}=(S_{n}^{\iota })$ with respect to $f$ with precision $\left\Vert \iota
\right\Vert $: $S_{n}^{\iota }(x)=i$ if $f^{n}(x)\in I_{i}^{d}$, $%
n=0,1,\ldots $ Then $\mathbf{S}^{\iota }$ is an ergodic stationary $N^{d}$%
-state random process or, equivalently, an ergodic source on $(I^{d},\left. 
\mathcal{B}\right\vert _{I^{d}},\mu )$ with finite alphabet $A^{\iota
}=\{1,...,N^{d}\}$ and output probability distribution $m=\mu \circ \phi
^{-1}$, with $\phi (x)=(S_{0}^{\iota }(x),S_{1}^{\iota }(x),...)\in A^{\iota 
\mathbb{N}}$, so that 
\begin{eqnarray}
&&\Pr \left( i_{0},\ldots ,i_{n-1}\right)   \notag \\
&=&\Pr \left( S_{0}^{\iota }=i_{0},\ldots ,S_{n-1}^{\iota }=i_{n-1}\right)  
\notag \\
&=&m\{s\in A^{\iota \mathbb{N}}:s_{0}=i_{0},...,s_{n-1}=i_{n-1}\}  \notag \\
&=&\mu \left( I_{i_{0}}^{d}\cap f^{-1}I_{i_{1}}^{d}\cap \ldots \cap
f^{-n+1}I_{i_{n-1}}^{d}\right) .  \label{Pr}
\end{eqnarray}%
In fact, $f$ and the left shift $\sigma $ on the sequences
$(S_{n}^{\iota }(x))$ are conjugate. A simple implementation of
$\mathbf{S}^{\iota }$ for $I=[0,1[$ and $N=10^{k}$ is the following:
$S_{n}^{\iota }(x)=\left\lfloor f^{n}(x)\cdot 10^{k}\right\rfloor
+1=\left\lceil f^{n}(x)\cdot 10^{k}\right\rceil $ with
$I_{i}=[(i-1)10^{-k},i10^{-k}[$ for $1\leq i\leq N$ . We see that
using simple observations as a finite alphabet measurement with
respect to $f$ provides a direct link between the entropies of
$\mathbf{S}^{\iota }$ and $f$. Accordingly, we define the
{\em permutation entropy rate of $f$} as
\begin{equation}
h_{\mu }^{\ast }(f):=\lim_{\left\Vert \iota \right\Vert \rightarrow
0}h_{m}^{\ast }(\mathbf{S}^{\iota })  \label{DefA}
\end{equation}
provided the limit exists. With this definition, and
Theorem~\ref{theorem:finite-alphabet}, we may prove the principal
result on ergodic dynamical systems.
\begin{thm} \label{theorem:dynamical-systems} If $f:I^{d}\rightarrow I^{d}$ 
is ergodic, then $h_{\mu}^{\ast }(f)=h_{\mu}(f)$. In words, the
permutation entropy rate of ergodic maps equals the metric entropy rate.
\end{thm}
\begin{pf}
If $h_{\mu }(f)=\infty $, the statement follows in
general (also for non-ergodic maps) from (\ref{Nonergodic}). If $h_{\mu
}(f)<\infty $, we have (see (\ref{Pr})) 
\begin{eqnarray*}
&&h_{m}\left( \mathbf{S}^{\iota }\right)  \\
&=&-\lim_{n\rightarrow \infty }\frac{1}{n}\sum \Pr \left( i_{0},\ldots
,i_{n-1}\right) \log \Pr \left( i_{0},\ldots ,i_{n-1}\right)  \\
&=&-\lim_{n\rightarrow \infty }\frac{1}{n}H_{\mu }\left(
\bigvee\limits_{i=0}^{n-1}f^{-i}\iota \right)  \\
&=&h_{\mu }(f,\iota ).
\end{eqnarray*}%
On the other hand, $h_{m}\left( \mathbf{S}^{\iota }\right)
=h_{m}^{\ast }\left( \mathbf{S}^{\iota }\right) $ by
Theorem~\ref{theorem:finite-alphabet} (since $\mathbf{S}^{\iota } $ is
ergodic with respect to the measure $m$).

Let $\gamma $ denote the finite generating partition of $f$ that, according
to Krieger's Theorem \cite{Walters}, must exist (due to $f$'s ergodicity and
finite metric entropy), so that $h_{\mu }(f)=h_{\mu }(f,\gamma )=h_{m}\left( 
\mathbf{S}^{\gamma }\right) $. We claim that 
\begin{equation*}
\lim_{\left\Vert \iota \right\Vert \rightarrow 0}h_{m}(\mathbf{S}^{\iota
})=h_{m}(\mathbf{S}^{\gamma })
\end{equation*}%
and, hence, 
\begin{equation*}
h_{\mu }^{\ast }(f)=\lim_{\left\Vert \iota \right\Vert \rightarrow
0}h_{m}^{\ast }(\mathbf{S}^{\iota })=\lim_{\left\Vert \iota \right\Vert
\rightarrow 0}h_{m}(\mathbf{S}^{\iota })=h_{\mu }(f).
\end{equation*}

\textit{Case 1}. Suppose that the elements of $\gamma $ are ($d$%
-dimensional) intervals or, more generally, that all elements of $\gamma $
consist of a finite number of intervals. In either case, taking if necessary
a refinement of $\gamma $ (thus, also a generator that we call $\gamma $ as
well) so that $\gamma $ becomes a product partition $\iota $ of $I^{d}$, we
deduce $h_{m}(\mathbf{S}^{\iota })=h_{m}(\mathbf{S}^{\gamma })=h_{\mu }(f)$
and the same is true for any further refinement of $\iota $.

\textit{Case 2}. If, otherwise, some component of $\gamma $ consists (modulo 
$0$) of infinitely many intervals, we can define a sequence of ever finer
partitions $(\iota _{n})_{n\in \mathbb{N}}$ of $I^{d}$ that, after an
hypothetical refinement can be assumed without restriction to be a product
partition (\textit{Case 1}) such that $\mathcal{A}(\iota _{n})$, the finite
sigma-algebras generated by the $\iota _{n}$, build an increasing sequence
and $\vee _{n=1}^{\infty }\mathcal{A}(\iota _{n})=\left. \mathcal{B}%
\right\vert _{I^{d}}$ ($\mathop{\rm mod}\nolimits0$). Then $h_{\mu
}(f)=\lim_{n\rightarrow \infty }h_{\mu }(f,\iota _{n})$ \cite{Walters}.

This proves our claim and the theorem.\qed
\end{pf}

\section{On the definition of permutation entropy rate for dynamical systems}

The original definition of Bandt, Keller and Pompe (BKP) \cite{Bandt} of the
permutation entropy of maps on intervals $I\subset \mathbb{R}$ involves
partitions of the form%
\begin{equation*}
P_{\pi }=\left\{ x\in I:f^{\pi (0)}(x)<f^{\pi (1)}(x)<\ldots <f^{\pi
(L-1)}(x)\right\} ,
\end{equation*}%
where $\pi \in \sigma _{L}$, here the set of permutations of $\{0,1,\ldots
,L-1\}$, $L\geq 2$. In fact, if $f$ is supposed to be piecewise monotone as
in \cite{Bandt} or just ergodic, as in our case, it is easy to show that%
\begin{equation*}
\mathcal{P}_{L}^{\ast }=\{P_{\pi }\neq \varnothing :\pi \in \sigma _{L}\}
\end{equation*}%
is a partition of $I$ (except maybe for a set of points of measure zero).
BKP define then the permutation entropy of order $L$ as 
\begin{eqnarray}
\bar{H}_{\mu }^{\ast \mathrm{BKP}}(f,L) &:=&\frac{1}{L-1}H_{\mu }(\mathcal{P}_{L}^{\ast })
\label{Hstar} \\
&=&-\frac{1}{L-1}\sum_{\pi \in \sigma _{L}}\mu (P_{\pi })\log \mu (P_{\pi })
\notag
\end{eqnarray}%
(compare to (\ref{Qpi})) and their permutation entropy rate of $f$ to be
\begin{equation}
h_{\mu }^{\ast \mathrm{BKP}}(f):=\lim_{L\rightarrow \infty }\bar{H}_{\mu }^{\ast \mathrm{BKP}}(f,L),  \label{DEfB}
\end{equation}%
provided the limit exists. They prove $h_{\mu }^{\ast
\mathrm{BKP}}(f)=h_{\mu }(f)$ for piecewise monotone maps on intervals
of $\mathbb{R}$, but in the more general case, ergodic maps it seems
that only the inequality $\lim \, \inf_{L\rightarrow \infty
}\bar{H}_{\mu }^{\ast }(f,L)\geq h_{\mu }(f)$---formally similar to
(\ref{Nonergodic})---can be proved, which we have done in Appendix C
for ergodic maps on $d$-dimensional intervals. Comparing such
particular results to the generality of Theorem~\ref{theorem:dynamical-systems}, we may conclude
that our definition (\ref{DefA}) of permutation entropy rate offers a
substantial advantage.

Note that the central distinction, which makes our formulation easier and
more natural, is that (\ref{DefA}) takes the limit of infinite long
conditioning ($L \rightarrow \infty)$ first, and the discrete limit ($\Delta
\rightarrow 0$) last, similarly to Kolmogorov-Sinai entropy rate, and as
opposed to~(\ref{DEfB}), where an explicit discretization was not taken. We
conjecture that for non-pathological dynamical systems of the sort one might
observe in Nature the two formulations are equivalent, but there are likely
to be some non-trivial technicalities involved in a rigorous analysis. For
example, \cite{Micha} shows a 1-dimensional map with an infinite number of
monotonicity intervals, where the \textit{topological} entropy rate and the
permutation version of the \textit{topological} entropy rate (i.e., counting
simply the number of distinct permutations with non-zero measure, and not
weighting them by their measure) are unequal: $h_{0}^{\ast
}(f)=\lim_{L\rightarrow \infty }\frac{1}{L-1}\log \left\vert \mathcal{P}%
_{L}^{\ast }\right\vert \neq h_{0}(f)$.

\section{\noindent Numerical examples and Discussion}

As a by-product of our result, the practitioner of time-series analysis will
find an alternative way to envision or, eventually, numerically estimate the
entropy rate of real sources. It is worth reminding that the entropy of
information sources can be measured by a variety of techniques that go
beyond counting word statistics and comprise different definitions of
`complexities' such as, for example, counting the patterns along a digital
(or digitalized) data sequence \cite{Lempel,Ziv2,Amigo}. Bandt and Pompe
refer, in \cite{Bandt2}, to the permutation entropy of time series as
complexity. That the entropy rate can also be computed by counting
permutations shows once again that it is a so general concept that can be
captured with different and seemingly blunt approaches.

\begin{figure}[tbp]
\includegraphics[width=\columnwidth]{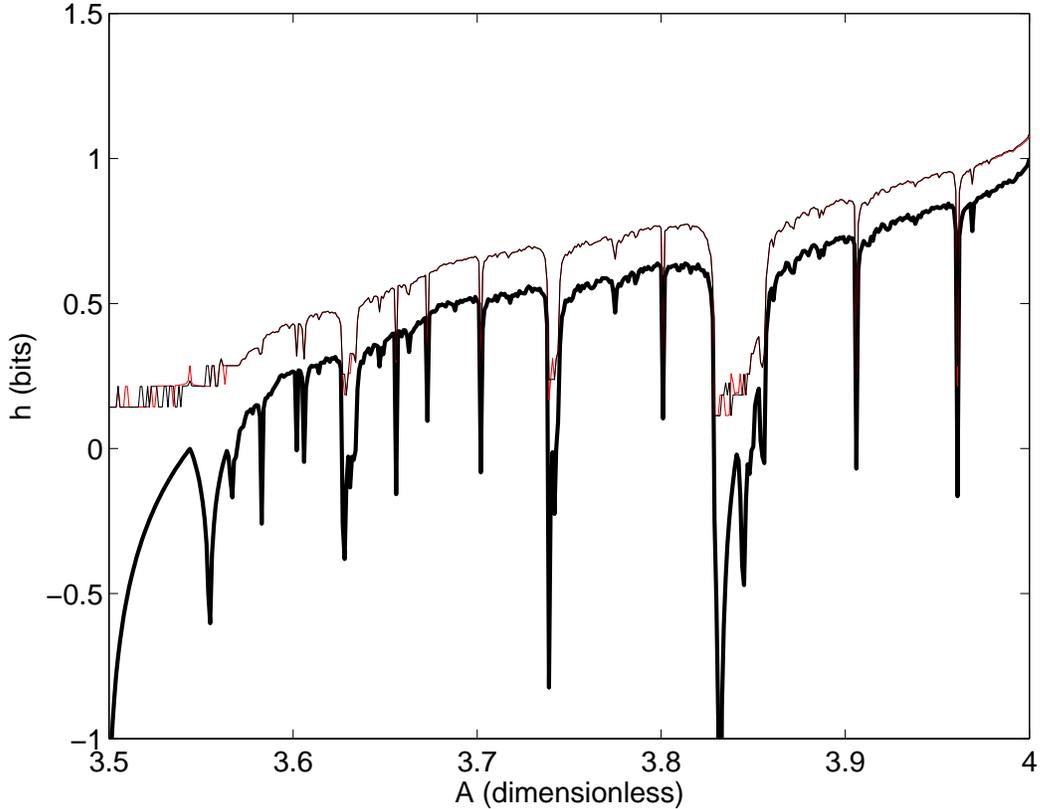}
\caption{(color online) Lyapunov exponent (black line, thick) of logistic
map and permutation entropy rate estimates $\hat{h} = \bar{H}^*({X}_1^{14})$
for $N=10^5, 10^6$ length time series from the map (red and black thin
lines). The permutation entropy estimate tracks changes in the Lyapunov
exponent (equal to the Kolmogorov-Sinai entropy rate where nonnegative)
well, with a nearly constant bias. Periodic orbits give a finite permutation
entropy, but the rate estimate would tend to zero given a sufficiently long
word.}
\end{figure}

\begin{figure}[tbp]
\includegraphics[width=\columnwidth]{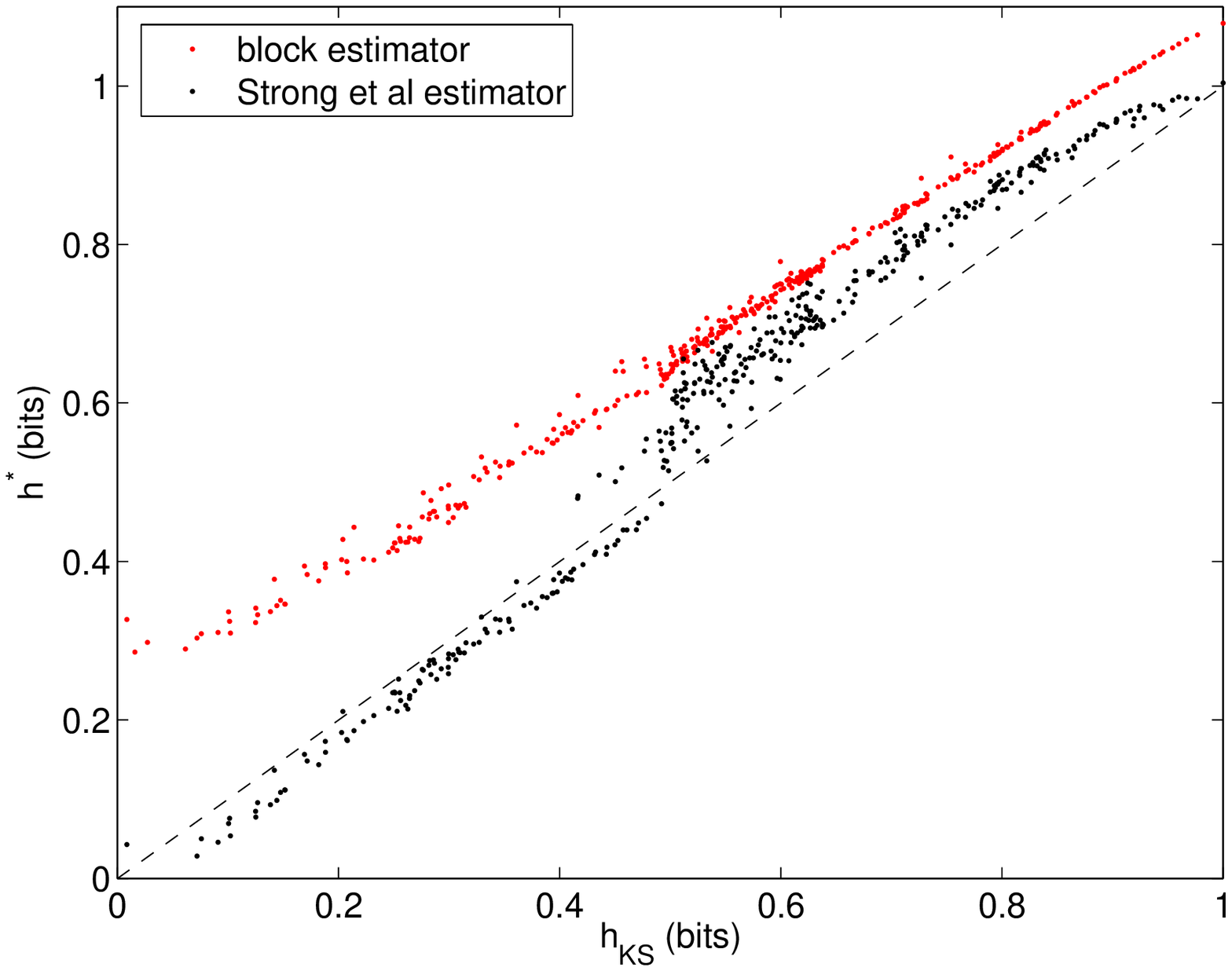}
\caption{(color online) Block entropy estimate (red points) at $L=14$ and
Strong et al~\protect\cite{Strong} fitted estimate (black points) as a
function of $h_{KS} = h_\protect\mu = \protect\lambda$ wherever $\protect%
\lambda \ge 0$. The scaling region Ansatz yields lower bias at cost of
increased variance. The block length and scaling region were chosen by hand,
a significant limitation.}
\end{figure}

We demonstrate numerical results on time series from the logistic map $%
x_{n+1}=Ax_{n}(1-x_{n})$. Figure 1 shows an estimate of the permutation
entropy rate estimate on noise-free data as a function of $A$, comparing the
Lyapunov exponent (computed from the orbit knowing the equation of motion)
to the permutation entropy. To be precise, we are estimating $h_{m}^{\ast }(%
\mathbf{S})$ with $\mathbf{S}$ discretized from the logistic map iterated at
the discretization of double-precision numerical representation, i.e., $%
\mathbf{S}$ is the output of a standard numerical iteration. The entropy
estimator of the block ranks was the plug-in estimator (substituting
observed frequencies for probabilities) plus the classical bias
correction, first order in $1/N$. The key unresolved issue in using
permutation entropies for empirical data analysis is, as with standard
Shannon entropy rate estimation, balancing the tension between larger
word lengths $L$, to capture more dependencies, and the loss of
sufficient sampling for good statistics in the ever larger discrete
space. The finite $L$ performance and convergence rate and bias of any
specific computational method are key issues when it comes to
accurately estimating the entropy rate of a source from observed
data. It is now appreciated that numerically estimating the Shannon
block entropy from finite data and, especially, the asymptotic entropy
rate, can be surprisingly tricky
\cite{Strong,Konto,Amigo,Kennel,Kennel2}. The theoretical definitions
of entropy rate do not necessarily lead to good statistical methods,
and superior alternatives have been developed over the many years
since Shannon. We believe that some of these ideas may similarly be
applicable to the permutation entropy situation.  Figure~2 shows a
very simple application of the part of the method of~\cite{Strong},
fitting an empirical asymptotic scaling $\bar{H}^{\ast
}(X_{1}^{L})=h_{L=\infty }+C/L$ for $L=13,14$, comparing to the block
estimate. This procedure shows a lower bias, but the specific choice
of scaling region $L$ (as with block entropy) is a key empirical
issue, and does not have a generally satisfactory resolution.

\begin{figure}[tbp]
\includegraphics[width=\columnwidth]{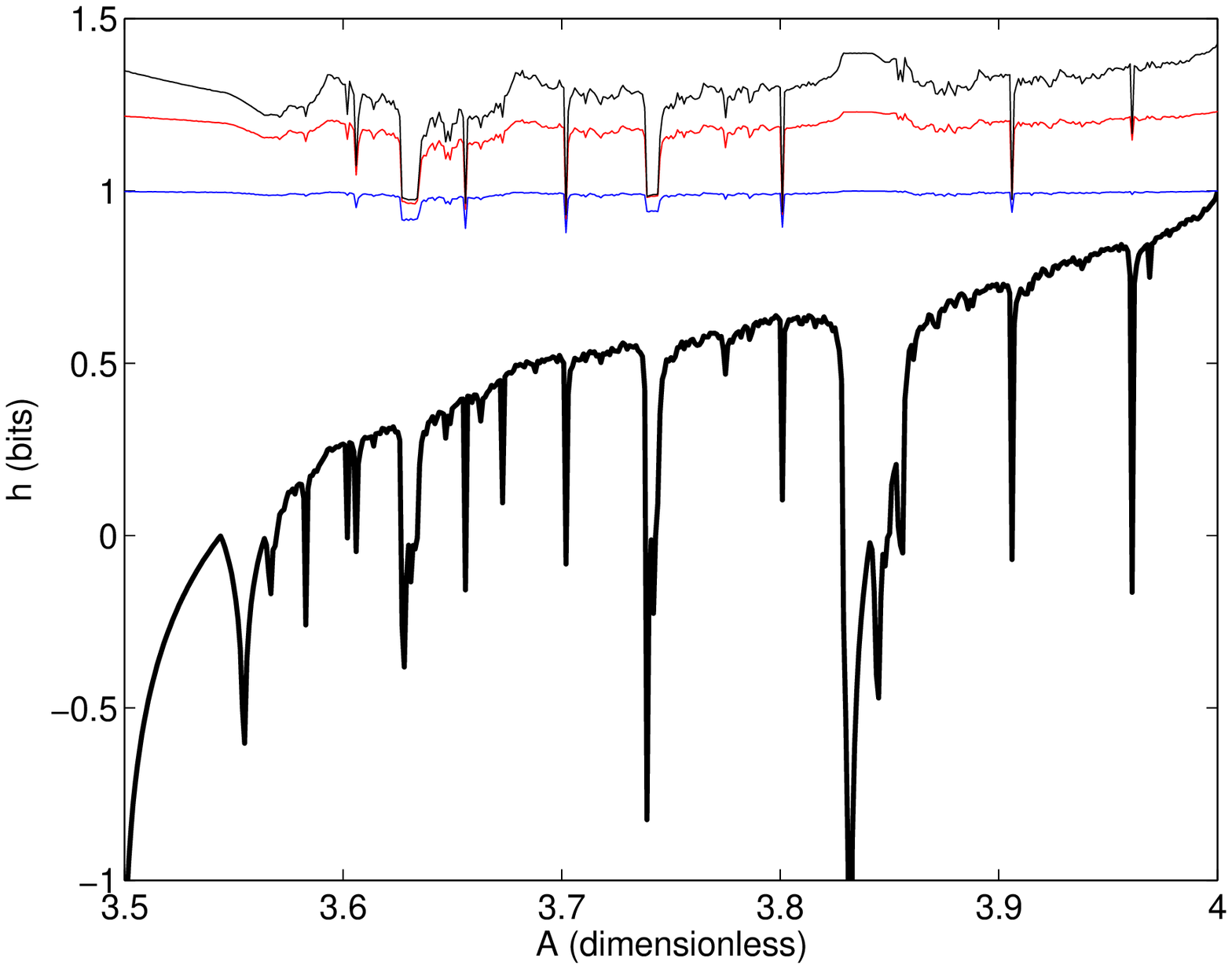}
\caption{(color online) Lyapunov exponent (black line, thick) of noise-free
logistic map and permutation entropy rate estimates $\hat{h} = \bar{H}^*({X}%
_1^{14})$ for $N=10^4,10^5, 10^6$ length time series from the map (blue, red
and black thin lines), contaminated with uniform zero-mean observational
noise of width 0.1. Here, the entropy of the underlying map is nearly
obliterated by the effect of the noise.}
\end{figure}

\begin{figure}[tbp]
\includegraphics[width=\columnwidth]{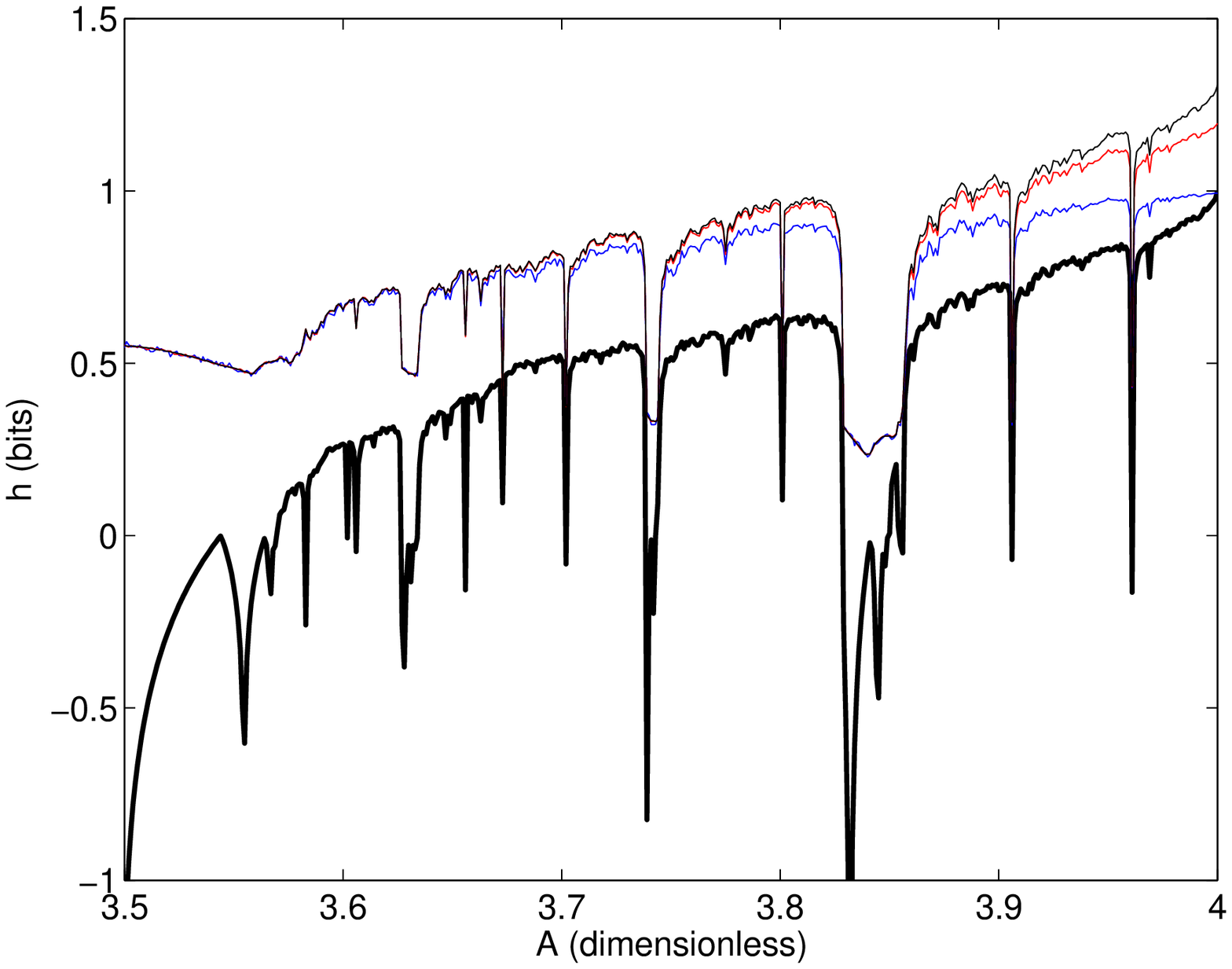}
\caption{(color online) Lyapunov exponent (black line, thick) of noise-free
logistic map and permutation entropy rate estimates $\hat{h} = \bar{H}^*({%
X^\Delta}_1^{14})$ for $N=10^4 10^5, 10^6$ length time series from the map
(blue, red and black thin lines), contaminated with uniform zero-mean
observational noise of width 0.1, and discretized to $\Delta = 0.2$. With
this discretization the entropy estimate tracks the macroscopic entropy from
the dynamics much better, though the bias is increased, as expected, since
the entropy due to noise still has some effect.}
\end{figure}

Also important for practical time-series analysis is the usual situation
where observations of a predominantly deterministic source is contaminated
with a small level of observational noise. Here, we recommend that the user 
\emph{fix} some discretization level $\Delta $ characteristic of the noise,
and evaluate the permutation entropies via entropies of rank words evaluated
from the discretized observables. Figure 3 shows analysis of permutations on
significantly noise-contaminated signals, with no explicit $\Delta $ (i.e.,
it is the size of the numerical precision of the computations). The
consequence is the permutation entropy is heavily dominated by the noise.
Figure~4 shows the restoration of monotonic scaling with $h_{\mu }$ when an
explicit, finite $\Delta =0.2$ is used to discretize the data before rank
variables are computed. Note that as computing ranks involves looking at the
difference between noise contaminated variables, when the characteristic
noise size is $0.1$, as in this example, an appropriate discretization scale
is $0.2$.

% The numerical investigation in~\cite{Bandt2} did not appear to do this (in effect evaluating $h^{**}$ at a small 
%$\Delta$ commensurate with the numerical resolution of the floating-point
%computations) with the consequence that when the time series was only
%observational noise the apparent short-term permutation entropy rate was
%quite large. In fact, it would not converge as the number of data increased
%(unlike for clean deterministic chaos) without the differential entropy
%correction. Without doing this discretization, the permutation entropy would
%end up being most sensitive to the complexity from the noise term, which is
%not usually what is desired.

For vector-valued sources, we applied lexicographic ordering and
construction of outer product variables in the proof. For analyzing chaotic
observed data, however, it may be acceptable to still use but one scalar
projection, subject to the traditional caveats of time-delay embedology. We
would expect that for appropriately mixing sources and generic observation
functions, the Kolmogorov-Sinai entropy estimated through that scalar still
equals the true value, and likewise so might permutation entropy rate. We
have found that numerically this appears to work in practice. With a direct
higher-dimensional product space, the undersampling issue becomes even more
difficult with increasing $L$, hence using scalars, as in a time-delay
embedding, may turn out to be a superior approach for observed time-series
of higher-dimensional sources.

\noindent \textbf{Acknowledgments.} We thank Domingo Morales (Universidad
Miguel Hern\'{a}ndez) for proof-reading some parts of the paper. J.M.A. was
partially supported by the Spanish Ministry of Education and Science, grant
GRUPOS 04/79. M.K. and L.K. thanks NSF for partial support.

\appendix

\section{Ergodic finite-alphabet information sources \label{sec:appendix1}}

\begin{pf*}{Proof of Lemma~\ref{lemma1}}
Given an ergodic information source $\mathbf{S}$,
\begin{equation*}
\lim_{k\rightarrow \infty }H_{m}(R_{k+1}^{k+l}|S_{1}^{k})=\lim_{k\rightarrow
\infty }H_{m}(S_{k+1}^{k+l}|S_{1}^{k})
\end{equation*}%
for all $l\geq 1$.
Consider $R_{k+1}=\sum_{i=1}^{k+1}\delta (S_{i}\leq
S_{k+1})$. For $a\in \{1,\ldots ,N\}$ define the \emph{sample frequency} of
the letter $a$ in the word $S_{1}^{k+1}$ to be 
\begin{equation*}
\vartheta _{k+1}(a)=\frac{1}{k+1}\sum_{i=1}^{k+1}\delta (S_{i}=a).
\end{equation*}%
With the help of $\vartheta _{k+1}(a)$ we may express $R_{k+1}$ in terms of $%
S_{i}$, $1\leq i\leq k+1$, namely, 
\begin{equation*}
R_{k+1}(S_{k+1})=(k+1)\sum_{a=1}^{S_{k+1}}\vartheta _{k+1}(a),
\end{equation*}%
where we assume the outcomes $S_{1}^{k+1}$ to be known. Then, the identity 
\begin{equation}
\Pr \left( R_{k+1}=y\right) =\sum_{q=1}^{N}\Pr \left( S_{k+1}=q\right)
\delta \left( R_{k+1}(q)=y\right)  \label{PrR}
\end{equation}%
give us the probability for observing some $R_{k+1}$ with value $y\in
\{1,\ldots ,k+1\}$ by means of $\Pr \left( S_{k+1}=q\right) $, $1\leq q\leq
N $. Since, given $S_{1}^{k}$, $R_{k+1}$ is a deterministic function of the
random variable $S_{k+1}$, i.e., $\Pr (R_{k+1}=y|S_{k+1}=q)=\delta
(R_{k+1}(q)=y)$, Eq. (\ref{PrR}) can be seen as an application of the law of
total probability.

Without loss of generality, we may first rearrange the sum in (\ref{PrR}) to
consider only those symbol values $q$ with non-zero $\Pr (S_{k+1}=q)$,
summing to $N^{\prime }\leq N$. Expand the sum, 
\begin{eqnarray*}
&&\Pr \left( R_{k+1}=y\right) = \\
&&\Pr \left( S_{k+1}=1\right) \delta \left[ y=(k+1)\vartheta _{k+1}(1)\right]
\\
&&+\Pr \left( S_{k+1}=2\right) \delta \left[ y=(k+1)(\vartheta
_{k+1}(1)+\vartheta _{k+1}(2))\right] \\
&&+\ldots +\Pr \left( S_{k+1}=N^{\prime }\right) \\
&&\times \delta \left[ y=(k+1)(\vartheta _{k+1}(1)+\ldots +\vartheta
_{k+1}(N^{\prime }))\right] .
\end{eqnarray*}%
Suppose all the relevant sample frequencies $\vartheta
_{k+1}(1),...,\vartheta _{k+1}(N^{\prime })$ are greater than zero. This
means that for any $y$, only a single one of the $\delta $-functions can be
nonzero, and hence we have a one-to-one transformation taking non-zero
elements from the distribution $\Pr (S_{k+1})$ without change into some bin
for $\Pr (R_{k+1})$. Since entropy is invariant to a renaming of the bins,
and the remaining zero probability bins add nothing to the entropy, we
conclude that, if $\vartheta _{k+1}(a)>0$ for all $a$ where the true
probability $\Pr (S_{k+1}=a)>0$ (i.e., $a=1,\ldots ,N^{\prime }$ after a
hypothetical rearrangement), then $%
H_{m}(R_{k+1}|S_{1}^{k})=H_{m}(S_{k+1}|S_{1}^{k})$. Because of the assumed
ergodicity, we can make the probability that $\vartheta _{k+1}(a)=0$ when $%
\Pr (S_{k+1}=a)>0$ to be arbitrarily small by taking $k$ to be sufficiently
large, and the claim follows for $l=1$.

This construction can be extended without change to words $S_{k+1}^{k+l}$ of
arbitrary length $l\geq 1$ via 
\begin{eqnarray*}
&&\Pr \left( R_{k+1}^{k+l}=y_{1}\ldots y_{l}\right) \\
&=&\sum_{q_{1},\ldots ,q_{l}=1}^{N^{\prime }}\left[ \Pr
(S_{k+1}^{k+l}=q_{1}\ldots q_{l})\times \delta (R_{k+1}(q_{1})=y_{1})\right.
\\
&&\left. ...\times \delta (R_{k+l}(q_{l})=y_{l})\right] .
\end{eqnarray*}%
Observe that if $\vartheta _{k+1}(a)>0$ for $1\leq a\leq N^{\prime }$, then
the same happens with $\vartheta _{k+2}(a)$,...,$\vartheta _{k+l}(a)$ and $%
H(R_{k+1}^{k+l}|S_{1}^{k})=H(S_{k+1}^{k+l}|S_{1}^{k})$ follows. Again,
ergodicity guarantees that there exist realizations $S_{1}^{k+l}$ whose
sample frequencies fulfill the said condition.\qed
\end{pf*}

As way of illustration, suppose that $S_{n}=0,1$ are independent random
variables with probability $\Pr (S_{n}=0)=\Pr (S_{n}=1)=\frac{1}{2}$. Given $%
S_{1}^{k}=s_{1}....s_{k}\in \{0,1\}^{k}$, set $N_{0}=\left\vert \{s_{i}=0%
\text{ in }S_{1}^{k}\}\right\vert $, $0\leq N_{0}\leq k$. Consider the case $%
L=2$ in Lemma 1. There are two possibilities:

(i) $0\leq N_{0}\leq k$. Then%
\begin{equation*}
\begin{array}{lll}
S_{k+1}^{k+2}=0,0 & \Rightarrow & R_{k+1}^{k+2}=N_{0}+1,N_{0}+2 \\ 
S_{k+1}^{k+2}=0,1 & \Rightarrow & R_{k+1}^{k+2}=N_{0}+1,k+2 \\ 
S_{k+1}^{k+2}=1,0 & \Rightarrow & R_{k+1}^{k+2}=k+2,N_{0}+1 \\ 
S_{k+1}^{k+2}=1,1 & \Rightarrow & R_{k+1}^{k+2}=k+1,k+2%
\end{array}%
\end{equation*}%
Each of these events has the joint probability%
\begin{equation*}
\Pr (N_{0}=\nu ,R_{k+1}^{k+2}=r_{k+1}^{k+2})=\frac{\binom{k}{\nu }}{2^{k}}%
\cdot \frac{1}{4}=\frac{1}{2^{k+2}}\binom{k}{\nu }
\end{equation*}%
and conditional probability%
\begin{equation*}
\Pr \left( R_{k+1}^{k+2}=r_{k+1}^{k+2}|N_{0}=\nu \right) =\frac{1}{4},
\end{equation*}%
where $0\leq \nu \leq k-1$ and $r_{k+1}^{k+2}=(\nu +1,\nu +2)$, $(\nu
+1,k+2) $, $(k+2,\nu +1)$ or $(k+1,k+2)$.

(ii) $N_{0}=k$. Then%
\begin{equation*}
\begin{array}{ll}
S_{k+1}^{k+2}=0,0\;\&\;S_{k+1}^{k+2}=0,1 & \&\;S_{k+1}^{k+2}=1,1 \\ 
& \Rightarrow R_{k+1}^{k+2}=k+1,k+2 \\ 
S_{k+1}^{k+2}=1,0 & \Rightarrow R_{k+1}^{k+2}=k+2,k+1%
\end{array}%
\end{equation*}%
These events have the joint probabilities%
\begin{eqnarray*}
\Pr \left( N_{0}=k,R_{k+1}^{k+2}=(k+1,k+2)\right) &=&\frac{1}{2^{k}}\cdot 
\frac{1}{4}\cdot 3=\frac{3}{2^{k+2}} \\
\Pr \left( N_{0}=k,R_{k+1}^{k+2}=(k+2,k+1)\right) &=&\frac{1}{2^{k}}\cdot 
\frac{1}{4}=\frac{1}{2^{k+2}}
\end{eqnarray*}%
and conditional probabilities%
\begin{eqnarray*}
\Pr \left( R_{k+1}^{k+2}=(k+1,k+2)|N_{0}=k\right) &=&\frac{3}{4} \\
\Pr \left( R_{k+1}^{k+2}=(k+1,k+2)|N_{0}=k\right) &=&\frac{1}{4}.
\end{eqnarray*}
From (i) and (ii), we get
\begin{eqnarray*}
&&H_{m}(R_{k+1}^{k+2}|S_{1}^{k}) \\
&=&-4\times \sum_{\nu =0}^{k-1}\frac{1}{2^{k+2}}\binom{k}{\nu }\log \frac{1}{
4}-\frac{3}{2^{k+2}}\log \frac{3}{4}-\frac{1}{2^{k+2}}\log \frac{1}{4} \\
&=&4\times \frac{2}{2^{k+2}}(2^{k}-1)+\frac{8}{2^{k+2}}-\frac{3}{2^{k+2}}
\log 3 \\
&=&2\left( 1-\frac{3}{2^{k+3}}\log 3\right) .
\end{eqnarray*}%
On the other hand, 
\begin{equation*}
H_{m}(S_{k+1}^{k+2}|S_{1}^{k})=H_{m}(S_{k+1}^{k+2})=2
\end{equation*}%
and so $H_{m}(R_{k+1}^{k+2}|S_{1}^{k})$ and $H_{m}(S_{k+1}^{k+2}|S_{1}^{k})$
are equal in the limit $k\rightarrow \infty $, as guaranteed by Lemma 1.

\section{Non-ergodic finite-alphabet sources}

In order to deal with the general, non-ergodic case, we appeal to the
theorem on ergodic decompositions \cite{Katok}: If $\Omega $ is a compact
metrizable space and $f:(\Omega ,\mathcal{F},\mu )\rightarrow (\Omega ,%
\mathcal{F},\mu )$ is continuous, then there is a partition of $\Omega $
into $f$-invariant subsets $\Omega _{w}$, each equipped with a sigma-algebra 
$\mathcal{F}_{w}$ and a probability measure $\mu _{w}$, such that $f$ acts
ergodically on each $(\Omega _{w},\mathcal{F}_{w},\mu _{w})$, the indexing
set being another probability space $(W,\mathcal{G},\nu )$ (in fact, a
Lebesgue space). Furthermore,%
\begin{equation*}
\mu (E)=\int_{W}\int_{E}d\mu _{w}d\nu (w)=\int_{W}\mu _{w}(E)d\nu
(w)\;\;\;(E\in \mathcal{F}).
\end{equation*}%
The family $\{\mu _{w}:w\in W\}$ is called the ergodic decomposition of $\mu
.$

If $\sigma $ is the shift on the (compact, metric) sequence space $(A^{%
\mathbb{N}},\mathcal{Z},m)$, the indexing set can be taken to be itself,
i.e.,%
\begin{equation}
m(C)=\int_{A^{\mathbb{N}}}\int_{C}dm_{s}dm(s)=\int_{A^{\mathbb{N}%
}}m_{s}(C)dm(s)\;\;\;(C\in \mathcal{Z}),  \label{decomp}
\end{equation}%
where $m_{\sigma (s)}=m_{s}$ \cite{Gray}. This result shows that any source
which is not ergodic can be represented as a mixture of ergodic subsources.
The next lemma states that such a decomposition holds also for the entropy
rate.

\begin{lem}[Ergodic Decomposition of the Entropy Rate~\cite{Gray}] 
\label{lemma:ergodic-decomposition}
Let $(A^{\mathbb{N}},\mathcal{Z},m,\sigma )$ be the
sequence space model of a stationary finite alphabet source $\mathbf{S}=(S_{n})$. Let
 $\{m_{s}:s\in A^{\mathbb{N}}\}$ be the ergodic decomposition of $m$. If
 $h_{m_{s}}(\mathbf{S})$ is $m$-integrable, then
\begin{equation}
h_{m}(\mathbf{S})=\int_{A^{\mathbb{N}}}h_{m_{s}}(\mathbf{S})dm(s).
\label{ErgodicDecomp}
\end{equation}
\end{lem}
\begin{thm} \label{theorem:non-ergodic} Under the assumptions of Lemma~\ref{lemma:ergodic-decomposition}, $\lim
\inf_{L\rightarrow \infty }\bar{H}_{m}^{\ast }(S_{1}^{L})\geq h_{m}(\mathbf{S})$ 
holds for any finite alphabet source $\mathbf{S}$.
\end{thm}
\begin{pf}
Fix $L\geq 2$. From (\ref{Qpi}) and (\ref{decomp}),
\begin{eqnarray}
&&\bar{H}_{m}^{\ast }(S_{1}^{L})  \notag \\
&=&-\frac{1}{L-1}\sum_{\pi \in \sigma _{L}}\left( \int_{A^{\mathbb{N}%
}}m_{s}(C_{\pi })dm(s)\right)  \notag \\
&&\times \log \left( \int_{A^{\mathbb{N}}}m_{s}(C_{\pi })dm(s)\right)  \notag
\\
&\geq &-\frac{1}{L-1}\sum_{\pi \in \sigma _{L}}\left( \int_{A^{\mathbb{N}%
}}m_{s}(C_{\pi })\log m_{s}(C_{\pi })dm(s)\right)  \label{Jensen} \\
&=&\int_{A^{\mathbb{N}}}\left( -\frac{1}{L-1}\sum_{\pi \in \sigma
_{L}}m_{s}(C_{\pi })\log m_{s}(C_{\pi })\right) dm(s)  \notag \\
&=&\int_{A^{\mathbb{N}}}\bar{H}_{m_{s}}^{\ast }(S_{1}^{L})dm(s),  \notag
\end{eqnarray}%
where in (\ref{Jensen}) we have used Jensen's inequality,%
\begin{equation*}
\Phi \left( \int_{A^{\mathbb{N}}}fd\mu \right) \leq \int_{A^{\mathbb{N}%
}}\Phi \circ fd\mu ,
\end{equation*}%
with $\Phi (t)=t\log t$ convex in $[0,\infty )$ and $f(s)=\mu _{s}(Q_{\pi
})\geq 0$.

Therefore,

\begin{eqnarray}
&&\lim_{L\rightarrow \infty }\inf \bar{H}_{m}^{\ast }(S_{1}^{L})  \notag \\
&\geq &\lim_{L\rightarrow \infty }\inf \int_{A^{\mathbb{N}}}\bar{H}%
_{m_{s}}^{\ast }(S_{1}^{L})dm(s)  \label{Fatou-2} \\
&\geq &\int_{A^{\mathbb{N}}}\left( \lim_{L\rightarrow \infty }\inf \bar{H}%
_{m_{s}}^{\ast }(S_{1}^{L})\right) dm(s)  \label{Fatou} \\
&=&\int_{A^{\mathbb{N}}}h_{m_{s}}^{\ast }(\mathbf{S})dm(s),  \label{Fatou+1}
\end{eqnarray}%
where we have applied Fatou's lemma in (\ref{Fatou}) to the sequence of
positive and $m$-measurable functions $\bar{H}_{m_{s}}^{\ast }(S_{1}^{L})$.
Observe that $h_{m_{s}}^{\ast }(\mathbf{S})$ exists for all $s\in A^{\mathbb{%
N}}$ (and is $m$-integrable as a function of $s$) since $h_{m_{s}}^{\ast }(%
\mathbf{S})=h_{m_{s}}(\mathbf{S})$ by Theorem 1 ($\mathbf{S}$ acts
ergodically on $(A_{s}^{\mathbb{N}},\mathcal{Z}_{s},m_{s})$). Therefore,%
\begin{equation*}
\lim_{L\rightarrow \infty }\inf \bar{H}_{m}^{\ast }(S_{1}^{L})\geq \int_{A^{%
\mathbb{N}}}h_{m_{s}}(\mathbf{S})dm(s)=h_{m}(\mathbf{S})
\end{equation*}%
by (\ref{ErgodicDecomp}).\qed
\end{pf} 

Theorem~\ref{theorem:non-ergodic} and Eqs. (\ref{Fatou-2}) and (\ref{Fatou+1}) yield:
\begin{cor}
If $h_{m}^{\ast }(\mathbf{S})=\lim_{L\rightarrow \infty
}\bar{H}_{m}^{\ast }(S_{1}^{L})$ exists for a non-ergodic
finite-alphabet source $\mathbf{S}$, then $h_{m}^{\ast
}(\mathbf{S})\geq h_{m}(\mathbf{S})$ and $h_{m}^{\ast
}(\mathbf{S})\geq \int_{A^{\mathbb{N}}}h_{m_{s}}^{\ast }(\mathbf{%
S})dm(s)$.
\end{cor}

\section{Interval maps}

Suppose first that $I$ is a one-dimensional interval and $f:I\rightarrow I$
an ergodic and $\mu $-preserving transformation, where $\mu $ is a measure
on $(I,\mathcal{B}\cap I),$ $\mathcal{B}$ being Borel sigma-algebra of $%
\mathbb{R}$.

\begin{lem} \label{lemma:interval-inequality}
If $f:I\rightarrow I$ is ergodic and $h_{\mu }(f)<\infty $, then 
$\lim \inf_{L\rightarrow \infty } \bar{H}_{\mu }^{\ast }(f,L)\geq h_{\mu }(f)$.
See (\ref{Hstar}) for the definition of $\bar{H}_{\mu }^{\ast
}(f,L)$. It follows, $h_{\mu }^{\ast \mathrm{BKP}}(f)\geq h_{\mu }(f)$.
\end{lem}
\begin{pf}
Let $\gamma $ be a finite generator of $f$
(Krieger's Theorem, \cite{Walters}). We split the proof in two parts. In the
first part we follow the approach of \cite[Sect. 3]{Bandt}.

\textit{Case 1}. Suppose that the elements of $\gamma $ are connected sets
(intervals) or, more generally, that all elements of $\gamma $ consist of a
finite number of intervals. In either case, taking if necessary a refinement
of $\gamma $ (thus, also a generator) that we call $\gamma $ as well, we
write without restriction $\gamma =\{I_{j},1\leq j\leq \left\vert \gamma
\right\vert \}$, were $I_{j}\subset I$ are intervals. This being the case,
let $c_{1}<c_{2}<...<c_{\left\vert \gamma \right\vert -1}$ be the points
that subdivide the interval $I=[a,b]$ into the $\left\vert \gamma
\right\vert $ intervals $I_{j}$ of the generator $\gamma $. We consider a
fixed $P_{\pi }\in \mathcal{P}_{L}^{\ast }$ and show that it can intersect
at most $(L+1)^{\left\vert \gamma \right\vert -1}$ sets of the partition $%
\gamma _{0}^{L-1}:=\vee _{i=0}^{L-1}f^{-i}(I_{j_{i}})$ with $%
I_{j_{0}},...,I_{j_{L-1}}\in \gamma $. For $x\in P_{\pi }$, let $\Delta
_{L}[x]$ denote the set in $\gamma _{0}^{L-1}$ that contains $x$. Thus, $%
\Delta _{L}[x]$ can be written as $I_{j_{0}}\cap f^{-1}(I_{j_{1}})\cap
...\cap f^{-(L-1)}(I_{j_{L-1}})$ with $I_{j_{0}},...,I_{j_{L-1}}\in \gamma $%
, so that it can be specified by the $n$-tuple $j[x]=(j_{0},...,j_{L-1})\in
\{1,...,\left\vert \gamma \right\vert \}^{L}$.

Now, $\pi $ is given by inequalities $x_{k_{1}}<...<x_{k_{L}}$ with $%
\{k_{1},...,k_{L}\}=\{0,...,L-1\}$ and $x_{k}=f^{k}(x)$. For each $x\in
P_{\pi }$ we can extend these inequalities so that they give the common
order of the $c_{r}$ and the $x_{k_{l}}$, where $1\leq r\leq \left\vert
\gamma \right\vert -1$ and $1\leq l\leq $ $L$. It follows that there are at
most $(L+1)^{\left\vert \gamma \right\vert -1}$ possible extended orders
since each $c_{r}$ has $L+1$ possible bins to go among the $x_{k_{l}}$ (as $%
x $ varies in $P_{\pi }$, the $L$ points $x_{k_{l}}$ defining the bins move
but do not cross each other). Moreover, when we know the common order of the 
$c_{r}$ and $x_{k_{l}}$, then $j[x]$ is uniquely determined (since $%
c_{j-1}<x_{k}<c_{j}$, implies $x_{k}\in I_{j}$ and thus $x\in f^{-k}(I_{j}),$
with $1\leq j\leq \left\vert \gamma \right\vert $, $c_{0}\equiv a$ and $%
c_{\left\vert \gamma \right\vert }\equiv b$).

Each $P_{\pi }\in \mathcal{P}_{L}^{\ast }$ is then the union of at most $%
(L+1)^{\left\vert \gamma \right\vert -1}$ sets $V_{k}\in $ $\gamma
_{0}^{L-1}\vee \mathcal{P}_{L}^{\ast }$ with total measure $\mu (P_{\pi })$.
Hence,%
\begin{eqnarray*}
&&-\sum_{k=1}^{(L+1)^{\left\vert \gamma \right\vert -1}}\mu (V_{k})\log \mu
(V_{k}) \\
&\leq &-\sum_{k=1}^{(L+1)^{\left\vert \gamma \right\vert -1}}\frac{\mu
(P_{\pi })}{(L+1)^{\left\vert \gamma \right\vert -1}}\log \frac{\mu (P_{\pi
})}{(L+1)^{\left\vert \gamma \right\vert -1}} \\
&=&-\mu (P_{\pi })\log \mu (P_{\pi })+(\left\vert \gamma \right\vert -1)\mu
(P_{\pi })\log (L+1)
\end{eqnarray*}%
and therefore, summing over all $\pi \in \sigma _{L}$,%
\begin{equation}
H_{\mu }(\gamma _{0}^{L-1})\leq H_{\mu }(\gamma _{0}^{L-1}\vee \mathcal{P}%
_{L}^{\ast })\leq H_{\mu }(\mathcal{P}_{L}^{\ast })+(\left\vert \gamma
\right\vert -1)\log (L+1).  \label{Case1}
\end{equation}%
It follows 
\begin{equation*}
\frac{1}{L-1}H_{\mu }(\mathcal{P}_{L}^{\ast })\geq \frac{1}{L-1}\left[
H_{\mu }(\gamma _{0}^{L-1})-(\left\vert \gamma \right\vert -1)\log (L+1)%
\right]
\end{equation*}%
and%
\begin{equation}
\lim_{L\rightarrow \infty }\inf \frac{1}{L-1}H_{\mu }(\mathcal{P}_{L}^{\ast
})\geq h_{\mu }(f)  \label{hmu(f)}
\end{equation}%
since $\gamma $ is a generator of $f$. Definition (\ref{Hstar}) completes
the proof in this case.

\textit{Case 2}. If some component of $\gamma $ consists of infinitely many
intervals, we can define a sequence of interval partitions $(\gamma
_{n})_{n\in \mathbb{N}}$ (\textit{Case 1}) such that $\mathcal{A}(\gamma
_{n})$, the finite sigma-algebras generated by the $\gamma _{n}$, build an
increasing sequence and $\vee _{n=1}^{\infty }\mathcal{A}(\gamma _{n})=%
\mathcal{B}$ ($\mathop{\rm mod}\nolimits0$). Then $h_{\mu
}(f)=\lim_{n\rightarrow \infty }h_{\mu }(f,\gamma _{n})$ \cite{Walters}.

We claim that, also in this case, Eq. (\ref{hmu(f)}) holds. Otherwise, for
every $\varepsilon >0$ and for every $L\geq 2$, there exists $L^{\prime }>L$
such that 
\begin{equation}
\frac{1}{L^{\prime }-1}H_{\mu }(\mathcal{P}_{L^{\prime }}^{\ast })<h_{\mu
}(f)-\varepsilon .  \label{h-eps}
\end{equation}%
Take now $n_{0}$ such that $\left\vert h_{\mu }(f)-h_{\mu }(f,\gamma
_{n})\right\vert <\varepsilon $ for all $n\geq n_{0}$. From (\ref{h-eps}) it
follows%
\begin{equation*}
\frac{1}{L^{\prime }-1}H_{\mu }(\mathcal{P}_{L^{\prime }}^{\ast })<h_{\mu
}(f,\gamma _{n_{0}})\leq \frac{1}{L^{\prime }-1}H((\gamma
_{n_{0}})_{0}^{L^{\prime }-1})
\end{equation*}%
because $\frac{1}{L}H((\gamma _{n_{0}})_{0}^{L-1})$ decreases monotonically
to $h_{\mu }(f,\gamma _{n_{0}})$. Use now (\ref{Case1}) to deduce%
\begin{eqnarray*}
&&\frac{1}{L^{\prime }-1}H_{\mu }(\mathcal{P}_{L^{\prime }}^{\ast }) \\
&<&h_{\mu }(f,\gamma _{n_{0}}) \\
&\leq &\frac{1}{L^{\prime }-1}H_{\mu }(\mathcal{P}_{L^{\prime }}^{\ast })+%
\frac{\left\vert \gamma _{n_{0}}\right\vert -1}{L^{\prime }-1}\log
(L^{^{\prime }}+1).
\end{eqnarray*}%
But the last term can be made arbitrarily small because the $L^{\prime }$
fulfilling (\ref{h-eps}) form an unbounded subsequence and $n_{0}$ is
independent of $L^{\prime }$. This contradiction proves our claim and
completes the proof. \qed
\end{pf}

More generally, let $I^{d}$ be now a proper, lexicographical ordered
interval of $\mathbb{R}^{d}$.
\begin{thm}
Let $f$ be an ergodic interval map in $\mathbb{R}^{d}$ fulfilling the above assumptions. If
 $h_{\mu }(f)<\infty $, then $\lim \inf_{L\rightarrow \infty }$ $\bar{H}%
_{\mu }^{\ast }(f,L)\geq h_{\mu }(f)$, where the permutation entropy
is defined by means of the product order of $\mathbb{R}^{d}$.
\end{thm}
\begin{pf*}{Proof outline}
As in Lemma~\ref{lemma:interval-inequality}, we split again its proof in
two cases. If (\textit{Case 1}) the generating partition is a product
partition or can be refined to a product partition%
\begin{equation*}
\gamma =\{I_{i}^{d},1\leq i\leq \left\vert \gamma \right\vert \},\;\;%
\overline{I_{i}^{d}}=[a_{1}^{(i)},b_{1}^{(i)}]\times ...\times \lbrack
a_{d}^{(i)},b_{d}^{(i)}],
\end{equation*}%
(whose elements are, without restriction, lexicographically ordered), then
the same approach used for one-dimensional intervals works through to Eq. (%
\ref{hmu(f)}). Otherwise (\textit{Case 2}), each element of $\gamma $ is the
countable union of disjoint intervals. They allow to define (after an
eventual refinement) a sequence of product partitions $(\gamma _{n})_{n\in 
\mathbb{N}}$ (\textit{Case 1}) such that $h_{\mu }(f)=\lim_{n\rightarrow
\infty }h_{\mu }(f,\gamma _{n}).$ The proof that $\lim \inf_{L\rightarrow
\infty }$ $\bar{H}_{\mu }^{\ast }(f,L)\geq h_{\mu }(f)$ is then completed
again by contradiction.\qed
\end{pf*}

\end{document}